\documentstyle[12pt,epsf]{article}

\renewcommand{\thefootnote}{\fnsymbol{footnote}}

\newcommand{\logqmms}{l_{qm}}

\newcommand{\logmsos}{L_{ms}}
\newcommand{\logmusos}{L_{s\mu}}
\newcommand{\logqmums}{l_{q\mu}}

\begin{document}    

\title{\vskip-3cm{\baselineskip14pt
\centerline{\normalsize\hfill MPI/PhT/97--012}
\centerline{\normalsize\hfill TTP97--11\footnote{The 
  complete postscript file of this
  preprint, including figures, is available via anonymous ftp at
  www-ttp.physik.uni-karlsruhe.de (129.13.102.139) as /ttp97-11/ttp97-11.ps 
  or via www at http://www-ttp.physik.uni-karlsruhe.de/cgi-bin/preprints.}}
\centerline{\normalsize\hfill hep-ph/9704222}
\centerline{\normalsize\hfill April 1997}
}
\vskip1.5cm
Mass Corrections to the Vector Current Correlator\footnote{
  Supported by BMBF under Contract 057KA92P, DFG under Contract Ku 502/8-1 
  and INTAS under Contract INTAS-93-744-ext.}
}
\author{
 K.G.~Chetyrkin$^{a,b}$,
 R.~Harlander$^{c}$\thanks{Supported by the ``Landesgraduiertenf\"orderung'' 
                           at the University of Karlsruhe.}, 
 J.H.~K\"uhn$^{c}$ 
 and
 M.~Steinhauser$^{a}$
}
\date{}
\maketitle

\begin{center}
$^a${\it Max-Planck-Institut f\"ur Physik,
    Werner-Heisenberg-Institut,\\ D-80805 Munich, Germany\\ }
  \vspace{3mm}
$^b${\it Institute for Nuclear Research, Russian Academy of Sciences,\\
   Moscow 117312, Russia\\}
  \vspace{3mm}
$^c${\it Institut f\"ur Theoretische Teilchenphysik,
    Universit\"at Karlsruhe,\\ D-76128 Karlsruhe, Germany.\\ }
\end{center}

\begin{abstract}
  \noindent Three-loop QCD corrections to the vector current correlator
  are considered.  The large momentum procedure is applied in order to
  evaluate mass corrections up to order $(m^2/q^2)^6$.  The inclusion of
  the first seven terms to the ratio
  $R=\sigma(e^+e^-\to\mbox{hadrons})/\sigma(e^+e^-\to\mu^+\mu^-)$ leads
  to reliable predictions from the high energy region down to relatively
  close to threshold.

\medskip
\noindent
PACS numbers: 12.38.-t, 12.38.Bx, 13.65.+i, 13.85.Lg.
\end{abstract}

\thispagestyle{empty}
\newpage
\setcounter{page}{1}


\renewcommand{\thefootnote}{\arabic{footnote}}
\setcounter{footnote}{0}


\section{Introduction}

One of the most precise measurements of the strong coupling constant
$\alpha_s$ is provided by the decay rate $\Gamma(Z\to\mbox{hadrons})$.
In the high energy limit the quark masses may often be neglected.
However, for precision measurements it is desirable to include also mass
corrections of the form $(m^2/s)^l$ with $l=1,2,3,\ldots$. This is
particularly valid if one wants to predict the total cross section for
$e^+e^-$ into hadrons in an energy region where $s$ and $m^2$ are of
comparable magnitude. In the massless limit corrections up to order
$\alpha_s^3$ are known
\cite{CheKatTka79DinSap79CelGon80,GorKatLar91SurSam91}.  Terms of order
$\alpha_s^2 m^2/s$, $\alpha_s^3 m^2/s$ 
\cite{GorKatLar86,CheKue90} and $\alpha_s^2 m^4/s^2$
\cite{CheKue94} are also available at present, providing an acceptable
approximation from the high energy region down to intermediate energy
values.  For the energy region closer to the production threshold terms
of higher order in $m^2/s$ are necessary.  In this paper a systematic
approach is presented to compute these terms. For the moment the vector
current correlator only is considered.  The starting point is thereby
the polarization function $\Pi(q^2)$.  With the method presented below
we are able to evaluate the three-loop terms up to order $(m^2/q^2)^6$ in
an expansion of $\Pi(q^2)$.

The polarization function $\Pi(q^2)$ is defined through
\begin{eqnarray}
\left(-g_{\mu\nu}q^2+q_\mu q_\nu\right)\,\Pi(q^2)
&=&i\int dx\,e^{iqx}\langle 0 |Tj_\mu(x) j_\nu(0)|0 \rangle
\end{eqnarray}
and the physical observable $R(s)$ is related to $\Pi(q^2)$ by
\begin{eqnarray}
R(s)&=&12\pi\,\mbox{Im}\,\Pi(q^2=s+i\epsilon).
\label{eqrtopiva}
\end{eqnarray}
It is convenient to write
\begin{eqnarray}
\Pi(q^2) &=& \Pi^{(0)}(q^2) 
         + \frac{\alpha_s(\mu^2)}{\pi} C_F \Pi^{(1)}(q^2)
         + \left(\frac{\alpha_s(\mu^2)}{\pi}\right)^2\Pi^{(2)}(q^2)
         + \ldots\,\,,
\\
\Pi^{(2)} &=&
                C_F^2       \Pi_A^{(2)}
              + C_A C_F     \Pi_{\it NA}^{(2)}
              + C_F T   n_l \Pi_l^{(2)}
              + C_F T       \Pi_F^{(2)},
\label{eqpi2}
\end{eqnarray}
and similarly for $R(s)$.  
The colour factors ($C_F=(N_c^2-1)/(2N_c)$ and $C_A=N_c$)
correspond to the Casimir operators of the fundamental
and adjoint representations, respectively.
For the numerical evaluation we set $N_c=3$.
The trace normalization of the fundamental representation is $T=1/2$.
The number of light (massless) quark flavours is denoted by $n_l$.
In Eq.~(\ref{eqpi2}) $\Pi_A^{(2)}$ is the abelian
contribution (quenched QED!) and $\Pi_{\it NA}^{(2)}$ is the non-abelian
part specific for QCD.  There are two fermionic contributions arising
from double-bubble diagrams: For $\Pi_l^{(2)}$ the quark in the inner
loop is massless, the external massive, whereas for $\Pi_F^{(2)}$ both
fermions have the same mass.  The result for $R_l^{(2)}(s)$ is known
analytically \cite{HoaKueTeu95} and will serve as check.  The case where
the external current couples to massless quarks and these via gluons to
massive ones is treated in \cite{HoaJezKueTeu94} and will not be
addressed here.

The outline of the paper is as follows: In Sect.~\ref{seclmp} we will
describe the procedure which allows the systematic expansion for
$\Pi(q^2)$ in analytic form for large external momentum and its
implementation in a program.  Sect.~\ref{secm12} contains the results
separated according to the colour factors.  In
Sect.~\ref{secim} the imaginary part of $\Pi(q^2)$ is considered and
compared to the result of a recent evaluation using
semi-analytical methods \cite{CheKueSte96}.  Finally the conclusions are
presented in Sect.~\ref{seccon}.


\section{\label{seclmp}Large momentum procedure}

In this section the basic rules of the
large momentum procedure are briefly described and their implementation
in programs is considered.

The methods which allow the expansion of Feynman diagrams 
containing either large masses (hard mass procedure)
or large external momenta (large momentum procedure)
have been used extensively in the recent past
\cite{Smi95}.
At one- and two-loop level it is in general still possible
to perform the diagrammatic expansion ``by hand'' and only
evaluate the integrals with algebraic programs. However,
at three loops this is almost impossible, in particular
if one is interested in an expansion up to high orders in $m^2/s$.

Following \cite{GorCheSmi} the prescriptions for the expansion of an
unrenormalized propagator type Feynman diagram in its large external momentum
read:
\begin{enumerate}
\item\label{presc1}
  Generate all subdiagrams of  the initial graph  such that
  \begin{enumerate}
  \item they contain both vertices through which  the  large 
    momentum enters and leaves the initial graph
  \item they become one-particle-irreducible when the two vertices are
    identified. 
  \end{enumerate}
\item\label{presc2}
  Taylor-expand the integrand of these subdiagrams in all small
  masses and external momenta generated by removing lines from the
  initial diagram. 
\item\label{presc3}
  In the initial diagram, shrink the subdiagram to a point and
  insert the result obtained from the expansion in \ref{presc2}.
\item\label{presc4}
  Sum over all terms.
\end{enumerate}
The subgraphs from step~\ref{presc1} are denoted {\it hard subgraphs}
or simply {\it subgraphs}, the reduced graphs, resulting from
step~\ref{presc3}, the {\it co-subgraphs}.  

\begin{figure}
  \begin{center}
  \begin{tabular}{c}
  \leavevmode
    \epsfxsize=10.cm
    \epsffile[115 455 450 700]{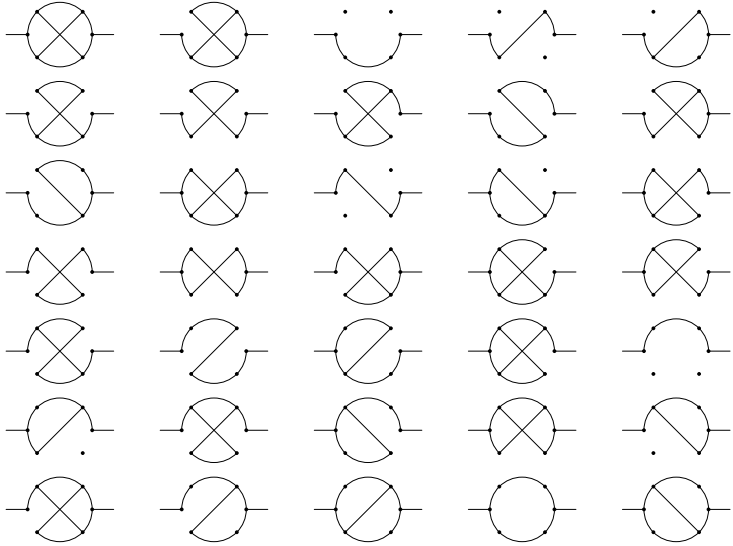}
  \end{tabular}
  \parbox{13.cm}{
    \caption[]{\label{no_exp_fig} Possible topologies for the subgraphs
      of an NO-type diagram.}
    }
  \end{center}
\end{figure}

For the non-planar three-loop topology all hard subgraphs are shown in
Fig.~\ref{no_exp_fig}. 
The corresponding co-subgraphs are obtained from
the full graph by shrinking the displayed lines to a point.
The number of  terms generated  by the large momentum procedure increases
rapidly with the number of loops in the initial diagram. Also, the
relation between the expansion momenta in the subgraph and the loop
momenta of the co-subgraph becomes non-trivial.
Nevertheless, the rules listed above provide an algorithm well suited
for the evaluation through a computer.
Restricting to three-loop two-point functions with large external
momentum and an arbitrary number of small masses, the procedure was
implemented using the language PERL.
The program can be divided into the following steps:
\begin{enumerate}
\item \label{perl_step}\begin{enumerate}
\item \label{gen_step} Generation of the relevant subgraphs and the
  corresponding co-subgraphs including the determination of their
  topologies, and
\item \label{dis_step} distribution of momenta in the subgraph and
  co-subgraph respecting the relations between them.
\end{enumerate}
\item \label{form_step} Calculation of these terms.
\end{enumerate}

In step \ref{gen_step}, one only considers topologies, disregarding any
properties of lines except their relative positions. Especially one
neglects their momenta, masses, particle types, etc.  For the
representation of a diagram we label its vertices by integers and
specify lines by their endpoints. A topology is thus described by the
collection of its lines.

To generate the subdiagrams 
we first note that the full graph is always one of
the hard subgraphs. The remaining ones are obtained by
going through the following steps:
\renewcommand{\labelenumi}{(\roman{enumi})}
\begin{enumerate}
\item Remove any combination of lines from the initial diagram.
\item Remove the emerging isolated dots and binary vertices.
\item Relabel the remaining vertices by 
  $\{1,\ldots,\#{\rm vertices}\}$ and build all permutations.
\item Compare with a table listing the basic one- and two-loop
  topologies, assign the proper topology if it matches one of the
  entries or, otherwise, discard the result.
\renewcommand{\labelenumi}{\Arabic{enumi}}
  
\end{enumerate}
If a subgraph passes the last step, the corresponding co-subgraph is
evidently fixed and its topology is determined in a similar way by
applying steps (ii)-(iv).  The
result of this procedure is a database containing all relevant sub- and
co-subgraphs, including information about their topology. Note that no
selection criteria related to line properties have been applied so far.
Thus, up to this point the procedure is universal and for each topology
this part of the program has to be run only once and for all.

In step \ref{dis_step}, the database resulting from \ref{gen_step} is
specialized to a specific diagram, including masses and momenta.  The
direction of the momentum carried by a line is connected with the order
of the labels representing this line.  Furthermore, for each subgraph
the small external momenta are routed in a way that they
touch as few lines as possible.  The output of step \ref{perl_step} is
then the input for step \ref{form_step}, together with all the necessary
administrative files like makefiles etc.

In step \ref{form_step} the two FORM \cite{form} packages MATAD and
MINCER are used which were already available and have been applied to
many different problems.  MINCER \cite{mincer} calculates massless
three-loop propagator type integrals and therefore applies to the hard
subgraphs.  MATAD calculates massive tadpole diagrams up to three loops
by using the corresponding recursion formulas obtained in \cite{Bro92}
and is used for the co-subgraphs accordingly.

Allowing only for one internal mass scale $m$, the final result is
then a power series in $m^2/q^2$, where $q$ is the (large) external
momentum. The coefficient functions contain 
numerical constants
together with $\ln^i(-q^2/m^2)$ and $\ln^j(-q^2/\mu^2)$
with $i\le 3$ and $j\le 2$, where $\mu$ is the renormalization scale.


\section{\label{secm12}Mass corrections to \boldmath{$\Pi(q^2)$}}

With the method presented in the previous section the first seven 
terms of the
expansion have been computed.  For the three-loop
case 18 initial diagrams have to be considered.  Altogether
240 subdiagrams are produced when the large momentum procedure is
applied. The corresponding numbers in the two- (one-) loop case are 2
(1) initial and 14 (3) subdiagrams.  The high energy approximation
for $\Pi(q^2)$ reads in the $\overline{\mbox{MS}}$ scheme ($\logqmms
\equiv \ln(-q^2/\bar{m}^2)$, $\logqmums \equiv \ln(-q^2/\mu^2)$):
\begin{eqnarray}
   \bar{\Pi}^{(0)} &=& {3 \over 16 \pi^2}\, \bigg\{
         {20\over 9}
          - {4\over 3}\,\logqmums
       + 8\,{\bar{m}^2\over q^2}
       + \left({\bar{m}^2\over q^2}\right)^{2} \, \left(
            4
          + 8\,\logqmms
          \right)
\nonumber\\&&\mbox{}
       + \left({\bar{m}^2\over q^2}\right)^{3} \, \left(
          - {64\over 9}
          + {32\over 3}\,\logqmms
          \right)
       + \left({\bar{m}^2\over q^2}\right)^{4} \, \left(
          - 22
          + 24\,\logqmms
          \right)
\nonumber\\&&\mbox{}
       + \left({\bar{m}^2\over q^2}\right)^{5} \, \left(
          - {992\over 15}
          + 64\,\logqmms
          \right)
       + \left({\bar{m}^2\over q^2}\right)^{6} \, \left(
          - {1852\over 9}
          + {560\over 3}\,\logqmms
          \right)
\bigg\} + \ldots\,\,,
\label{pi0bar}
\\[.4cm]
   \bar{\Pi}^{(1)} &=& {3 \over 16 \pi^2}\, \bigg\{
         {55\over 12}
          - 4\,\zeta_3
          - \logqmums
       + {\bar{m}^2\over q^2} \, \left(
            16
          - 12\,\logqmums
          \right)
\nonumber\\&&\mbox{}
       + \left({\bar{m}^2\over q^2}\right)^{2} \, \bigg[
            {2\over 3}
          + 16\,\zeta_3
          + 22\,\logqmms
          - 24\,\logqmms\,\logqmums
          + 12\,\logqmms^2
          \bigg]
\nonumber\\&&\mbox{}
       + \left({\bar{m}^2\over q^2}\right)^{3} \, \bigg[
          - {736\over 27}
          + {1184\over 27}\,\logqmms
          + {200\over 9}\,\logqmms^2
          + \left( 48
          - 48\,\logqmms \right)\,\logqmums
          \bigg]
\nonumber\\&&\mbox{}
       + \left({\bar{m}^2\over q^2}\right)^{4} \, \bigg[
          - {40313\over 216}
          + {1199\over 9}\,\logqmms
          + {229\over 3}\,\logqmms^2
          + \left( 168
          - 144\,\logqmms \right)\,\logqmums
          \bigg]
\nonumber\\&&\mbox{}
       + \left({\bar{m}^2\over q^2}\right)^{5} \, \bigg[
          - {2641601\over 3375}
          + {279196\over 675}\,\logqmms
          + {12248\over 45}\,\logqmms^2
          + \left( 592
          - 480\,\logqmms \right)\,\logqmums
          \bigg]
\nonumber\\&&\mbox{}
       + \left({\bar{m}^2\over q^2}\right)^{6} \, \bigg[
          - {6117262\over 2025}
          + {184066\over 135}\,\logqmms
          + {8992\over 9}\,\logqmms^2
\nonumber\\&&\mbox{\hspace{1.0cm}}
          + \left( 2132
          - 1680\,\logqmms \right)\,\logqmums
          \bigg]
\bigg\} + \ldots\,\,,
\label{pi1bar}
\\[.4cm]
   \bar{\Pi}^{(2)}_{A} &=& {3 \over 16 \pi^2}\, \bigg\{
       - {143\over 72}
          - {37\over 6}\,\zeta_3
          + 10\,\zeta_5
          + {1\over 8}\,\logqmums
\nonumber\\&&\mbox{}
       + {\bar{m}^2\over q^2} \, \bigg[
            {1667\over 24}
          - {5\over 3}\,\zeta_3
          - {70\over 3}\,\zeta_5
          - {51\over 2}\,\logqmums
          + 9\,\logqmums^2
          \bigg]
\nonumber\\&&\mbox{}
       + \left({\bar{m}^2\over q^2}\right)^{2} \, \bigg[
            {127\over 2}
          + 96\,\zeta_3
          - 48\,\zeta_4
          - 20\,\zeta_5
          + 8\,B_4
          + \left( 36
          + 24\,\zeta_3 \right)\,\logqmms
          + {33\over 2}\,\logqmms^2
\nonumber\\&&\mbox{\hspace{1.0cm}}
          + 12\,\logqmms^3
          + \left( 31
          - 48\,\zeta_3
          - 33\,\logqmms
          - 36\,\logqmms^2 \right)\,\logqmums
          + \left( - 18
          + 36\,\logqmms \right)\,\logqmums^2
          \bigg]
\nonumber\\&&\mbox{}
       + \left({\bar{m}^2\over q^2}\right)^{3} \, \bigg[
          - {177163\over 2916}
          - 13\,\zeta_3
          - 64\,\zeta_4
          + {320\over 3}\,\zeta_5
          + {32\over 3}\,B_4
\nonumber\\&&\mbox{\hspace{1.0cm}}
          + \left( {9988\over 81}
          + {248\over 3}\,\zeta_3 \right)\,\logqmms
          + {239\over 3}\,\logqmms^2
          + {2896\over 81}\,\logqmms^3
\nonumber\\&&\mbox{\hspace{1.0cm}}
          + \left( {1750\over 9}
          - {410\over 3}\,\logqmms
          - 100\,\logqmms^2 \right)\,\logqmums
          + \left( - 144
          + 108\,\logqmms \right)\,\logqmums^2
          \bigg]
\nonumber\\&&\mbox{}
       + \left({\bar{m}^2\over q^2}\right)^{4} \, \bigg[
          - {12656479\over 15552}
          - {14293\over 54}\,\zeta_3
          - 144\,\zeta_4
          + {880\over 3}\,\zeta_5
          + 24\,B_4
\nonumber\\&&\mbox{\hspace{1.0cm}}
          + \left( {26783\over 108}
          + {826\over 3}\,\zeta_3 \right)\,\logqmms
          + {494059\over 1296}\,\logqmms^2
          + {53701\over 324}\,\logqmms^3
\nonumber\\&&\mbox{\hspace{1.0cm}}
          + \left( {48263\over 36}
          - {1765\over 3}\,\logqmms
          - 458\,\logqmms^2 \right)\,\logqmums
          + \left( - 612
          + 432\,\logqmms \right)\,\logqmums^2
          \bigg]
\nonumber\\&&\mbox{}
       + \left({\bar{m}^2\over q^2}\right)^{5} \, \bigg[
          - {539215192817\over 116640000}
          - {1056859\over 900}\,\zeta_3
          - 384\,\zeta_4
          + {3200\over 3}\,\zeta_5
          + 64\,B_4
\nonumber\\&&\mbox{\hspace{1.0cm}}
          + \left( {1191314219\over 3240000}
          + {14234\over 15}\,\zeta_3 \right)\,\logqmms
          + {90647953\over 54000}\,\logqmms^2
          + {2940547\over 4050}\,\logqmms^3
\nonumber\\&&\mbox{\hspace{1.0cm}}
          + \left( {328233\over 50}
          - {105554\over 45}\,\logqmms
          - {6124\over 3}\,\logqmms^2 \right)\,\logqmums
          + \left( - 2580
          + 1800\,\logqmms \right)\,\logqmums^2
          \bigg]
\nonumber\\&&\mbox{}
       + \left({\bar{m}^2\over q^2}\right)^{6} \, \bigg[
          - {420607059143\over 19440000}
          - {13059229\over 2700}\,\zeta_3
          - 1120\,\zeta_4
          + 4000\,\zeta_5
          + {560\over 3}\,B_4
\nonumber\\&&\mbox{\hspace{1.0cm}}
          + \left( {133099291\over 972000}
          + {16842\over 5}\,\zeta_3 \right)\,\logqmms
          + {54076013\over 7200}\,\logqmms^2
          + {8575579\over 2700}\,\logqmms^3
\nonumber\\&&\mbox{\hspace{1.0cm}}
          + \left( {13274779\over 450}
          - {142256\over 15}\,\logqmms
          - 8992\,\logqmms^2 \right)\,\logqmums
\nonumber\\&&\mbox{\hspace{1.0cm}}
          + \left( - 10854
          + 7560\,\logqmms \right)\,\logqmums^2
          \bigg]
\bigg\} + \ldots\,\,,
\label{pi2abar}
\\[.4cm]
   \bar{\Pi}^{(2)}_{\it NA} &=& {3 \over 16 \pi^2}\, \bigg\{
         {44215\over 2592}
          - {227\over 18}\,\zeta_3
          - {5\over 3}\,\zeta_5
          + \left( - {41\over 8}
          + {11\over 3}\,\zeta_3 \right)\,\logqmums
          + {11\over 24}\,\logqmums^2
\nonumber\\&&\mbox{}
       + {\bar{m}^2\over q^2} \, \bigg[
            {1447\over 24}
          + {16\over 3}\,\zeta_3
          - {85\over 3}\,\zeta_5
          - {185\over 6}\,\logqmums
          + {11\over 2}\,\logqmums^2
          \bigg]
\nonumber\\&&\mbox{}
       + \left({\bar{m}^2\over q^2}\right)^{2} \, \bigg[
          - {5051\over 216}
          + {529\over 9}\,\zeta_3
          + 24\,\zeta_4
          - 30\,\zeta_5
          - 4\,B_4
          + \left( {2123\over 36}
          - 12\,\zeta_3 \right)\,\logqmms
\nonumber\\&&\mbox{\hspace{1.0cm}}
          + {76\over 3}\,\logqmms^2
          + {11\over 3}\,\logqmms^3
          + \left( - {11\over 18}
          - {44\over 3}\,\zeta_3
          - {105\over 2}\,\logqmms
          - 11\,\logqmms^2 \right)\,\logqmums
          + 11\,\logqmms\,\logqmums^2
          \bigg]
\nonumber\\&&\mbox{}
       + \left({\bar{m}^2\over q^2}\right)^{3} \, \bigg[
          - {1278391\over 11664}
          + {149\over 2}\,\zeta_3
          + 32\,\zeta_4
          - {80\over 3}\,\zeta_5
          - {16\over 3}\,B_4
\nonumber\\&&\mbox{\hspace{1.0cm}}
          + \left( {247627\over 1944}
          - {158\over 3}\,\zeta_3 \right)\,\logqmms
          + {745\over 18}\,\logqmms^2
          + {352\over 81}\,\logqmms^3
\nonumber\\&&\mbox{\hspace{1.0cm}}
          + \left( {7262\over 81}
          - {8494\over 81}\,\logqmms
          - {550\over 27}\,\logqmms^2 \right)\,\logqmums
          + \left( - 22
          + 22\,\logqmms \right)\,\logqmums^2
          \bigg]
\nonumber\\&&\mbox{}
       + \left({\bar{m}^2\over q^2}\right)^{4} \, \bigg[
          - {13377061\over 31104}
          + {4840\over 27}\,\zeta_3
          + 72\,\zeta_4
          - {220\over 3}\,\zeta_5
          - 12\,B_4
\nonumber\\&&\mbox{\hspace{1.0cm}}
          + \left( {1154479\over 2592}
          - {512\over 3}\,\zeta_3 \right)\,\logqmms
          + {98381\over 648}\,\logqmms^2
          + {973\over 81}\,\logqmms^3
\nonumber\\&&\mbox{\hspace{1.0cm}}
        + \left( {1030099\over 2592}
        - {34141\over 108}\,\logqmms
        - {2519\over 36}\,\logqmms^2 \right)\,\logqmums
        + \left( - 77
        + 66\,\logqmms \right)\,\logqmums^2
        \bigg]
\nonumber\\&&\mbox{}
       + \left({\bar{m}^2\over q^2}\right)^{5} \, \bigg[
          - {418731949411\over 233280000}
          + {335497\over 600}\,\zeta_3
          + 192\,\zeta_4
          - {800\over 3}\,\zeta_5
          - 32\,B_4
\nonumber\\&&\mbox{\hspace{1.0cm}}
          + \left( {5920594861\over 3888000}
          - {2863\over 5}\,\zeta_3 \right)\,\logqmms
          + {1433459\over 2400}\,\logqmms^2
          + {14063\over 324}\,\logqmms^3
\nonumber\\&&\mbox{\hspace{1.0cm}}
       + \left( {61358611\over 40500}
       - {2077289\over 2025}\,\logqmms
       - {33682\over 135}\,\logqmms^2 \right)\,\logqmums
          + \left( - {814\over 3}
          + 220\,\logqmms \right)\,\logqmums^2
          \bigg]
\nonumber\\&&\mbox{}
       + \left({\bar{m}^2\over q^2}\right)^{6} \, \bigg[
          - {34153293329\over 4665600}
          + {902921\over 450}\,\zeta_3
          + 560\,\zeta_4
          - 1000\,\zeta_5
          - {280\over 3}\,B_4
\nonumber\\&&\mbox{\hspace{1.0cm}}
          + \left( {49904336431\over 9720000}
          - 1988\,\zeta_3 \right)\,\logqmms
          + {50112223\over 21600}\,\logqmms^2
          + {1376567\over 8100}\,\logqmms^3
\nonumber\\&&\mbox{\hspace{1.0cm}}
        + \left( {34271558\over 6075}
        - {2845663\over 810}\,\logqmms
        - {24728\over 27}\,\logqmms^2 \right)\,\logqmums
\nonumber\\&&\mbox{\hspace{1.0cm}}
          + \left( - {5863\over 6}
          + 770\,\logqmms \right) \,\logqmums^2
          \bigg]
\bigg\} + \ldots\,\,,
\label{pi2nabar}
\\[.4cm]
   \bar{\Pi}^{(2)}_{l} &=& {3 \over 16 \pi^2}\, \bigg\{
       - {3701\over 648}
          + {38\over 9}\,\zeta_3
          + \left( {11\over 6}
          - {4\over 3}\,\zeta_3 \right)\,\logqmums
          - {1\over 6}\,\logqmums^2
\nonumber\\&&\mbox{}
       + {\bar{m}^2\over q^2} \, \bigg[
          - {95\over 6}
          + {26\over 3}\,\logqmums
          - 2\,\logqmums^2
          \bigg]
\nonumber\\&&\mbox{}
       + \left({\bar{m}^2\over q^2}\right)^{2} \, \bigg[
          - {83\over 54}
          - {224\over 9}\,\zeta_3
          - {145\over 9}\,\logqmms
          - {20\over 3}\,\logqmms^2
          - {4\over 3}\,\logqmms^3
\nonumber\\&&\mbox{\hspace{1.0cm}}
          + \left( {2\over 9}
          + {16\over 3}\,\zeta_3
          + 14\,\logqmms
          + 4\,\logqmms^2 \right)\,\logqmums
          - 4\,\logqmms\,\logqmums^2
          \bigg]
\nonumber\\&&\mbox{}
       + \left({\bar{m}^2\over q^2}\right)^{3} \, \bigg[
            {1292\over 729}
          - {128\over 9}\,\zeta_3
          - {11182\over 243}\,\logqmms
          - {344\over 27}\,\logqmms^2
          - {128\over 81}\,\logqmms^3
\nonumber\\&&\mbox{\hspace{1.0cm}}
          + \left( - {1816\over 81}
          + {2264\over 81}\,\logqmms
          + {200\over 27}\,\logqmms^2 \right)\,\logqmums
          + \left( 8
          - 8\,\logqmms \right)\,\logqmums^2
          \bigg]
\nonumber\\&&\mbox{}
       + \left({\bar{m}^2\over q^2}\right)^{4} \, \bigg[
            {11941\over 144}
          - 8\,\zeta_3
          - {42641\over 324}\,\logqmms
          - {4883\over 108}\,\logqmms^2
          - {127\over 27}\,\logqmms^3
\nonumber\\&&\mbox{\hspace{1.0cm}}
          + \left( - {70553\over 648}
          + {2279\over 27}\,\logqmms
          + {229\over 9}\,\logqmms^2 \right)\,\logqmums
          + \left( 28
          - 24\,\logqmms \right)\,\logqmums^2
          \bigg]
\nonumber\\&&\mbox{}
       + \left({\bar{m}^2\over q^2}\right)^{5} \, \bigg[
            {391086983\over 911250}
          + {448\over 45}\,\zeta_3
          - {2461553\over 6075}\,\logqmms
          - {113648\over 675}\,\logqmms^2
          - {6488\over 405}\,\logqmms^3
\nonumber\\&&\mbox{\hspace{1.0cm}}
       + \left( - {4306601\over 10125}
       + {549196\over 2025}\,\logqmms
       + {12248\over 135}\,\logqmms^2 \right)\,\logqmums
       + \left( {296\over 3}
       - 80\,\logqmms \right)\,\logqmums^2
       \bigg]
\nonumber\\&&\mbox{}
       + \left({\bar{m}^2\over q^2}\right)^{6} \, \bigg[
            {165619333\over 91125}
          + {2432\over 27}\,\zeta_3
          - {1763021\over 1350}\,\logqmms
          - {51451\over 81}\,\logqmms^2
          - {4664\over 81}\,\logqmms^3
\nonumber\\&&\mbox{\hspace{1.0cm}}
        + \left( - {9715012\over 6075}
        + {373066\over 405}\,\logqmms
        + {8992\over 27}\,\logqmms^2 \right)\,\logqmums
\nonumber\\&&\mbox{\hspace{1.0cm}}
          + \left( {1066\over 3}
          - 280\,\logqmms \right)\,\logqmums^2
          \bigg]
\bigg\} + \ldots\,\,,
\label{pi2lbar}
\\[.4cm]
   \bar{\Pi}^{(2)}_{F} &=& {3 \over 16 \pi^2}\, \bigg\{
       - {3701\over 648}
          + {38\over 9}\,\zeta_3
          + \left( {11\over 6}
          - {4\over 3}\,\zeta_3 \right)\,\logqmums
          - {1\over 6}\,\logqmums^2
\nonumber\\&&\mbox{}
       + {\bar{m}^2\over q^2} \, \bigg[
          - {223\over 6}
          + 16\,\zeta_3
          + {26\over 3}\,\logqmums
          - 2\,\logqmums^2
          \bigg]
\nonumber\\&&\mbox{}
       + \left({\bar{m}^2\over q^2}\right)^{2} \, \bigg[
          - {1505\over 54}
          + {352\over 9}\,\zeta_3
          + \left( - {439\over 9}
          + 8\,\zeta_3 \right)\,\logqmms
          - {23\over 3}\,\logqmms^2
          - {4\over 3}\,\logqmms^3
\nonumber\\&&\mbox{\hspace{1.0cm}}
          + \left( {2\over 9}
          + {16\over 3}\,\zeta_3
          + 14\,\logqmms
          + 4\,\logqmms^2 \right)\,\logqmums
          - 4\,\logqmms\,\logqmums^2
          \bigg]
\nonumber\\&&\mbox{}
       + \left({\bar{m}^2\over q^2}\right)^{3} \, \bigg[
            {69374\over 729}
          + {1120\over 27}\,\zeta_3
          - {6050\over 81}\,\logqmms
          - {436\over 81}\,\logqmms^2
          - {112\over 81}\,\logqmms^3
\nonumber\\&&\mbox{\hspace{1.0cm}}
          + \left( - {1816\over 81}
          + {2264\over 81}\,\logqmms
          + {200\over 27}\,\logqmms^2 \right)\,\logqmums
          + \left( 8
          - 8\,\logqmms \right)\,\logqmums^2
          \bigg]
\nonumber\\&&\mbox{}
       + \left({\bar{m}^2\over q^2}\right)^{4} \, \bigg[
            {175813\over 432}
          + 186\,\zeta_3
          - {87737\over 324}\,\logqmms
          - {8501\over 108}\,\logqmms^2
          - {370\over 27}\,\logqmms^3
\nonumber\\&&\mbox{\hspace{1.0cm}}
          + \left( - {70553\over 648}
          + {2279\over 27}\,\logqmms 
          + {229\over 9}\,\logqmms^2
\right)\,\logqmums
          + \left( 28
          - 24\,\logqmms \right)\,\logqmums^2
          \bigg]
\nonumber\\&&\mbox{}
       + \left({\bar{m}^2\over q^2}\right)^{5} \, \bigg[
            {843731341\over 455625}
          + {84064\over 135}\,\zeta_3
          - {5236781\over 10125}\,\logqmms
          - {646424\over 2025}\,\logqmms^2
          - {8072\over 81}\,\logqmms^3
\nonumber\\&&\mbox{\hspace{1.0cm}}
       + \left( - {4306601\over 10125}
       + {549196\over 2025}\,\logqmms
       + {12248\over 135}\,\logqmms^2 \right)\,\logqmums
       + \left( {296\over 3}
       - 80\,\logqmms \right)\,\logqmums^2
       \bigg]
\nonumber\\&&\mbox{}
       + \left({\bar{m}^2\over q^2}\right)^{6} \, \bigg[
            {2624352259\over 364500}
          + {41212\over 27}\,\zeta_3
          + {3852533\over 24300}\,\logqmms
          - {34499\over 54}\,\logqmms^2
          - {18142\over 27}\,\logqmms^3
\nonumber\\&&\mbox{\hspace{1.0cm}}
        + \left( - {9715012\over 6075}
        + {373066\over 405}\,\logqmms
        + {8992\over 27}\,\logqmms^2 \right)\,\logqmums
\nonumber\\&&\mbox{\hspace{1.0cm}}        + \left( {1066\over 3}
        - 280\,\logqmms \right)\,\logqmums^2
        \bigg]
\bigg\} + \ldots\,\,,
\label{pi2fbar}
\end{eqnarray}
where $\bar{m}$ is the $\overline{\mbox{MS}}$ renormalized mass at scale
$\mu^2$ and $\zeta$ is Riemann's zeta-function with the values
$\zeta_2=\pi^2/6$, $\zeta_3\approx1.20206$, $\zeta_4=\pi^4/90$ and
$\zeta_5\approx1.03693$.  $B_4$ is a constant typical for massive
three-loop integrals with 
$B_4 = -{13\over 2} \zeta_4 - 4\zeta_2 \ln^2 2 + {2\over 3}\ln^4 2 
+ 16 {\rm Li}_4 ({1\over 2})
\approx-1.762800$ \cite{Bro92}.  
The overall renormalization of $\bar{\Pi}(q^2)$ is also performed in the
$\overline{\mbox{MS}}$ scheme, i.e.~in the expressions obtained after
renormalization of $m$ and $\alpha_s$ only the poles are subtracted.
The expressions renormalized in the conventional QED scheme are obtained
by subtracting $\bar{\Pi}(0)$ given e.g.~in 
\cite{Bro92,CheKueSte96}
in order to obtain $\Pi(0) = 0$.
The $(m^2/q^2)^2$ terms in the case of QED can also be found in
\cite{CheHarKueSte96}.


\section{\label{secim}\boldmath{$\sigma(e^+e^-\to {\rm hadrons})$}}

According to Eq.~(\ref{eqrtopiva}) the ratio $R(s)$ is obtained by taking the
imaginary part arising from the $\ln(-q^2)$ terms of the above results.
Note that starting from the quartic term logarithms in the mass
$\bar{m}$ appear which cannot be removed by a choice of the
renormalization scale $\mu$.  A closer look into the method used for the
calculation shows that starting from this order massive tadpoles appear,
these being the source for such logarithms in agreement with the general
discussion of \cite{BroGen84,CheSpi87}.  
For $\mu^2=s$ which is the natural scale
at high energies, the $(\bar{m}^2/s)^0$ and $(\bar{m}^2/s)^1$ terms are
free of logarithms.
We refrain from listing the corresponding results which are trivially
obtained from Eqs.(\ref{pi0bar})-(\ref{pi2fbar}).

Using the relation between the $\overline{\mbox{MS}}$ and the
on-shell mass \cite{GraBroGraSch90}
leads to the ratio $R(s)$ expressed in terms of the pole mass
($\logmsos \equiv \ln(m^2/s), \logmusos \equiv \ln(s/\mu^2)$):
\begin{eqnarray}
   R^{(0)} &=& 3\, \bigg\{
         1
       - 6 \, \left({m^2\over s}\right)^{2}
       - 8 \, \left({m^2\over s}\right)^{3}
       - 18 \, \left({m^2\over s}\right)^{4}
       - 48 \, \left({m^2\over s}\right)^{5}
\nonumber\\&&\mbox{\hspace{0.8cm}}
       - 140 \, \left({m^2\over s}\right)^{6}
\label{r0os}
\bigg\} + \ldots\,\,,\\[.4cm]
   R^{(1)} &=& 3\, \bigg\{
         {3\over 4}
       + 9 \, {m^2\over s}
       + \left({m^2\over s}\right)^{2} \, \bigg[
            {15\over 2}
          - 18 \logmsos
          \bigg]
       + \left({m^2\over s}\right)^{3} \, \bigg[
          - {188\over 9}
          - {116\over 3} \logmsos
          \bigg]
\nonumber\\&&\mbox{}
       + \left({m^2\over s}\right)^{4} \, \bigg[
          - {983\over 12}
          - {203\over 2} \logmsos
          \bigg]
       + \left({m^2\over s}\right)^{5} \, \bigg[
          - {61699\over 225}
          - {4676\over 15} \logmsos
          \bigg]
\nonumber\\&&\mbox{}
       + \left({m^2\over s}\right)^{6} \, \bigg[
          - {84743\over 90}
          - {3064\over 3} \logmsos
          \bigg]
\label{r1os}
\bigg\} + \ldots\,\,,\\[.4cm]
   R^{(2)}_{A} &=& 3\, \bigg\{
       - {3\over 32}
       + {m^2\over s}\, \bigg[
            {9\over 8}
          + {27\over 2} \logmsos
          \bigg]
\nonumber\\&&\mbox{}
       + \left({m^2\over s}\right)^{2} \, \bigg[
          - {345\over 16}
          + \left( 99
          - 72\,\ln 2\right)\,\zeta_2
          + 36\,\zeta_3
          + {81\over 4} \logmsos
          - 27 \logmsos^2
          \bigg]
\nonumber\\&&\mbox{}
       + \left({m^2\over s}\right)^{3} \, \bigg[
            {12469\over 216}
          + \left( {2582\over 9}
          - 144\,\ln 2\right)\,\zeta_2
          - 26\,\zeta_3
          - {139\over 2} \logmsos
          - {886\over 9} \logmsos^2
          \bigg]
\nonumber\\&&\mbox{}
       + \left({m^2\over s}\right)^{4} \, \bigg[
            {6551\over 72}
          + \left( {64717\over 72}
          - 432\,\ln 2\right)\,\zeta_2
          - {197\over 2}\,\zeta_3
          - {360005\over 864} \logmsos
\nonumber\\&&\mbox{\hspace{0.8cm}}
          - {45277\over 144} \logmsos^2
          \bigg]
\nonumber\\&&\mbox{}
       + \left({m^2\over s}\right)^{5} \, \bigg[
            {259771381\over 4320000}
          + \left( {2773147\over 900}
          - 1440\,\ln 2\right)\,\zeta_2
          - {3517\over 10}\,\zeta_3
\nonumber\\&&\mbox{\hspace{0.8cm}}
          - {63112847\over 36000} \logmsos
          - {1963147\over 1800} \logmsos^2
          \bigg]
\nonumber\\&&\mbox{}
       + \left({m^2\over s}\right)^{6} \, \bigg[
          - {748405531\over 1296000}
          + \left( {6599179\over 600}
          - 5040\,\ln 2\right)\,\zeta_2
          - {12663\over 10}\,\zeta_3
\nonumber\\&&\mbox{\hspace{0.8cm}}
          - {33526867\over 4800} \logmsos
          - {4709179\over 1200} \logmsos^2
          \bigg]
\label{r2aos}
\bigg\} + \ldots\,\,,\\[.4cm]
   R^{(2)}_{\it NA} &=& 3\, \bigg\{
         {123\over 32}
          - {11\over 4}\,\zeta_3
          - {11\over 16} \logmusos
       + {m^2\over s}\, \bigg[
            {185\over 8}
          - {33\over 4} \logmusos
          \bigg]
\nonumber\\&&\mbox{}
       + \left({m^2\over s}\right)^{2} \, \bigg[
            {77\over 3}
          + \left( {9\over 2}
          + 36\,\ln 2\right)\,\zeta_2
          + 11\,\zeta_3
\nonumber\\&&\mbox{\hspace{0.8cm}}
          - {381\over 8} \logmsos
          + {33\over 4} \logmsos^2
          + \left( - {55\over 8}
          + {33\over 2} \logmsos\right) \logmusos
          \bigg]
\nonumber\\&&\mbox{}
       + \left({m^2\over s}\right)^{3} \, \bigg[
          - {61951\over 2592}
          + \left( {26\over 9}
          + 72\,\ln 2\right)\,\zeta_2
          + {43\over 2}\,\zeta_3
\nonumber\\&&\mbox{\hspace{0.8cm}}
          - {11779\over 108} \logmsos
          + 22 \logmsos^2
          + \left( {517\over 27}
          + {319\over 9} \logmsos\right) \logmusos
          \bigg]
\nonumber\\&&\mbox{}
       + \left({m^2\over s}\right)^{4} \, \bigg[
          - {372361\over 1728}
          + \left( - {2579\over 72}
          + 216\,\ln 2\right)\,\zeta_2
          + 74\,\zeta_3
\nonumber\\&&\mbox{\hspace{0.8cm}}
          - {61961\over 216} \logmsos
          + {10793\over 144} \logmsos^2
          + \left( {10813\over 144}
          + {2233\over 24} \logmsos\right) \logmusos
          \bigg]
\nonumber\\&&\mbox{}
       + \left({m^2\over s}\right)^{5} \, \bigg[
          - {4611741517\over 5184000}
          + \left( - {63869\over 360}
          + 720\,\ln 2\right)\,\zeta_2
          + {4989\over 20}\,\zeta_3
\nonumber\\&&\mbox{\hspace{0.8cm}}
          - {34493231\over 43200} \logmsos
          + {20357\over 80} \logmsos^2
          + \left( {678689\over 2700}
          + {12859\over 45} \logmsos\right) \logmusos
          \bigg]
\nonumber\\&&\mbox{}
       + \left({m^2\over s}\right)^{6} \, \bigg[
          - {13915043077\over 4320000}
          + \left( - {1316833\over 1800}
          + 2520\,\ln 2\right)\,\zeta_2
          + 861\,\zeta_3
\nonumber\\&&\mbox{\hspace{0.8cm}}
          - {103349851\over 43200} \logmsos
          + {1058411\over 1200} \logmsos^2
          + \left( {932173\over 1080}
          + {8426\over 9} \logmsos\right) \logmusos
          \bigg]
\bigg\} + \ldots\,\,,
\nonumber\\         
\label{r2naos}
\\[.4cm]
   R^{(2)}_{l} &=& 3\, \bigg\{
       - {11\over 8}
          + \zeta_3
          + {1\over 4} \logmusos
       + {m^2\over s}\, \bigg[
          - {13\over 2}
          + 3 \logmusos
          \bigg]
\nonumber\\&&\mbox{}
       + \left({m^2\over s}\right)^{2} \, \bigg[
          - {35\over 6}
          - 18\,\zeta_2
          - 4\,\zeta_3
          + {27\over 2} \logmsos
          - 3 \logmsos^2
          + \left( {5\over 2}
          - 6 \logmsos\right) \logmusos
          \bigg]
\nonumber\\&&\mbox{}
       + \left({m^2\over s}\right)^{3} \, \bigg[
            {1282\over 81}
          - {304\over 9}\,\zeta_2
          + {752\over 27} \logmsos
          - 8 \logmsos^2
          + \left( - {188\over 27}
          - {116\over 9} \logmsos\right) \logmusos
          \bigg]
\nonumber\\&&\mbox{}
       + \left({m^2\over s}\right)^{4} \, \bigg[
            {7091\over 96}
          - {260\over 3}\,\zeta_2
          + {5291\over 72} \logmsos
          - {53\over 2} \logmsos^2
          + \left( - {983\over 36}
          - {203\over 6} \logmsos\right) \logmusos
          \bigg]
\nonumber\\&&\mbox{}
       + \left({m^2\over s}\right)^{5} \, \bigg[
            {2712517\over 10125}
          - {11872\over 45}\,\zeta_2
\nonumber\\&&\mbox{\hspace{0.8cm}}
          + {142327\over 675} \logmsos
          - 92 \logmsos^2
          + \left( - {61699\over 675}
          - {4676\over 45} \logmsos\right) \logmusos
          \bigg]
\nonumber\\&&\mbox{}
       + \left({m^2\over s}\right)^{6} \, \bigg[
            {15168713\over 16200}
          - {7744\over 9}\,\zeta_2
\nonumber\\&&\mbox{\hspace{0.8cm}}
          + {9721\over 15} \logmsos
          - {2972\over 9} \logmsos^2
          + \left( - {84743\over 270}
          - {3064\over 9} \logmsos\right) \logmusos
          \bigg]
\label{r2los}
\bigg\} + \ldots\,\,,\\[.4cm]
   R^{(2)}_{F} &=& 3\, \bigg\{
       - {11\over 8}
          + \zeta_3
          + {1\over 4} \logmusos
       + {m^2\over s}\, \bigg[
          - {13\over 2}
          + 3 \logmusos
          \bigg]
\nonumber\\&&\mbox{}
       + \left({m^2\over s}\right)^{2} \, \bigg[
            {2\over 3}
          + 18\,\zeta_2
          - 10\,\zeta_3
          + 12 \logmsos
          - 3 \logmsos^2
          + \left( {5\over 2}
          - 6 \logmsos\right) \logmusos
          \bigg]
\nonumber\\&&\mbox{}
       + \left({m^2\over s}\right)^{3} \, \bigg[
            {4\over 3}
          + {352\over 9}\,\zeta_2
          + {350\over 9} \logmsos
          - {76\over 9} \logmsos^2
          + \left( - {188\over 27}
          - {116\over 9} \logmsos\right) \logmusos
          \bigg]
\nonumber\\&&\mbox{}
       + \left({m^2\over s}\right)^{4} \, \bigg[
            {20233\over 288}
          + {533\over 6}\,\zeta_2
          + {1673\over 72} \logmsos
          - {25\over 4} \logmsos^2
\nonumber\\&&\mbox{\hspace{0.8cm}}
          + \left( - {983\over 36}
          - {203\over 6} \logmsos\right) \logmusos
          \bigg]
\nonumber\\&&\mbox{}
       + \left({m^2\over s}\right)^{5} \, \bigg[
          - {54559\over 6750}
          + {3592\over 45}\,\zeta_2
          - {1157\over 75} \logmsos
          + {4328\over 45} \logmsos^2
\nonumber\\&&\mbox{\hspace{0.8cm}}
          + \left( - {61699\over 675}
          - {4676\over 45} \logmsos\right) \logmusos
          \bigg]
\nonumber\\&&\mbox{}
       + \left({m^2\over s}\right)^{6} \, \bigg[
          - {9214697\over 6480}
          - 1105\,\zeta_2
          + {346981\over 540} \logmsos
          + {18937\over 18} \logmsos^2
\nonumber\\&&\mbox{\hspace{0.8cm}}
          + \left( - {84743\over 270}
          - {3064\over 9} \logmsos\right) \logmusos
          \bigg]
\label{r2fos}
\bigg\} + \ldots\,\,.
\end{eqnarray}
The quartic terms are in agreement with the results of \cite{CheKue94}. 
For the contribution containing massless fermions, $R_l^{(2)}$,
a comparison with the full analytical result is possible
\cite{HoaKueTeu95}. We found complete agreement up to the
order considered.

\begin{figure}
\begin{center}
\begin{tabular}{cc}
    \leavevmode
    \epsfxsize=5.5cm
    \epsffile[110 265 465 560]{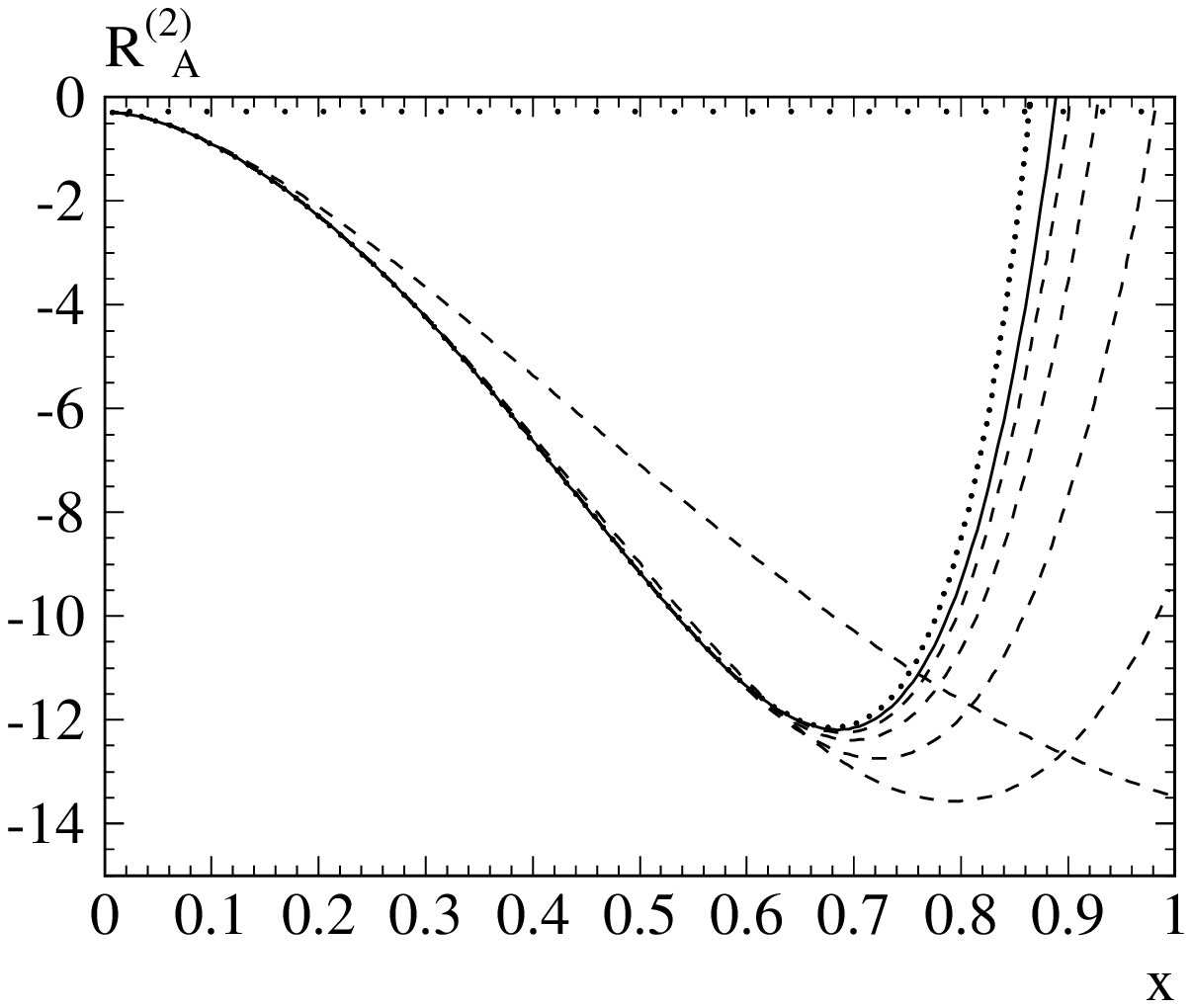}&
    \epsfxsize=5.5cm
    \epsffile[110 265 465 560]{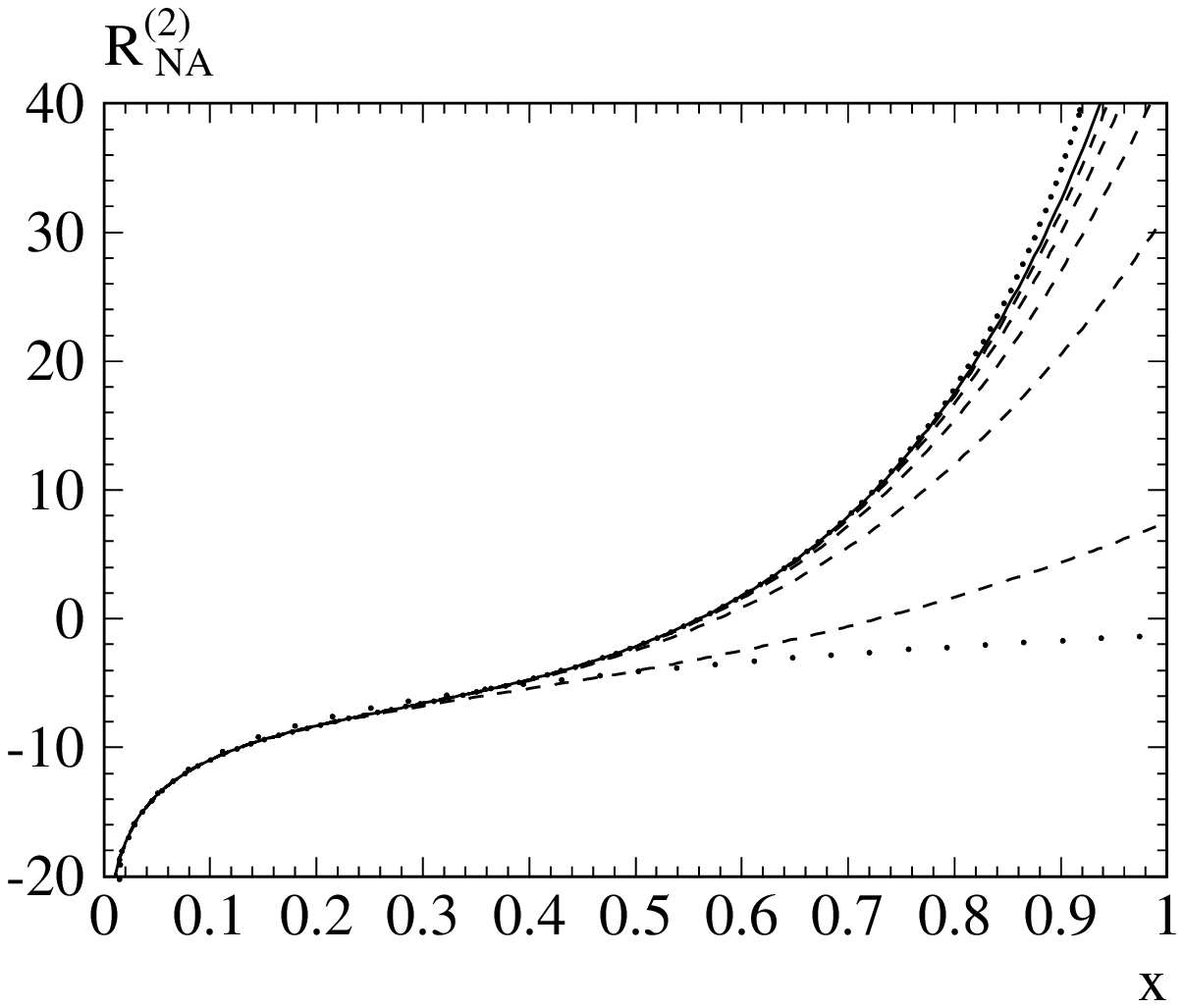}\\
    \epsfxsize=5.5cm
    \epsffile[110 265 465 560]{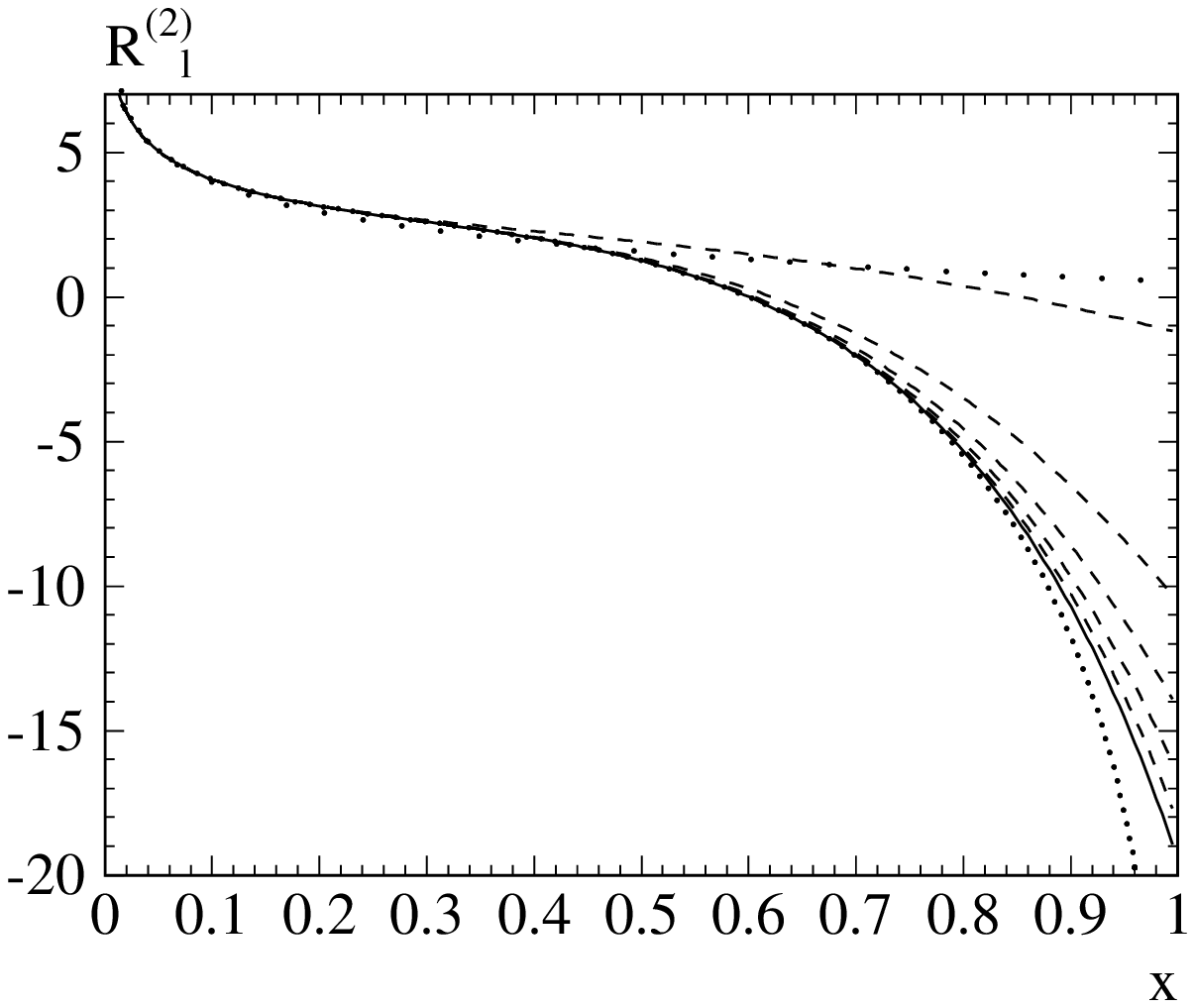}&
    \epsfxsize=5.5cm
    \epsffile[110 265 465 560]{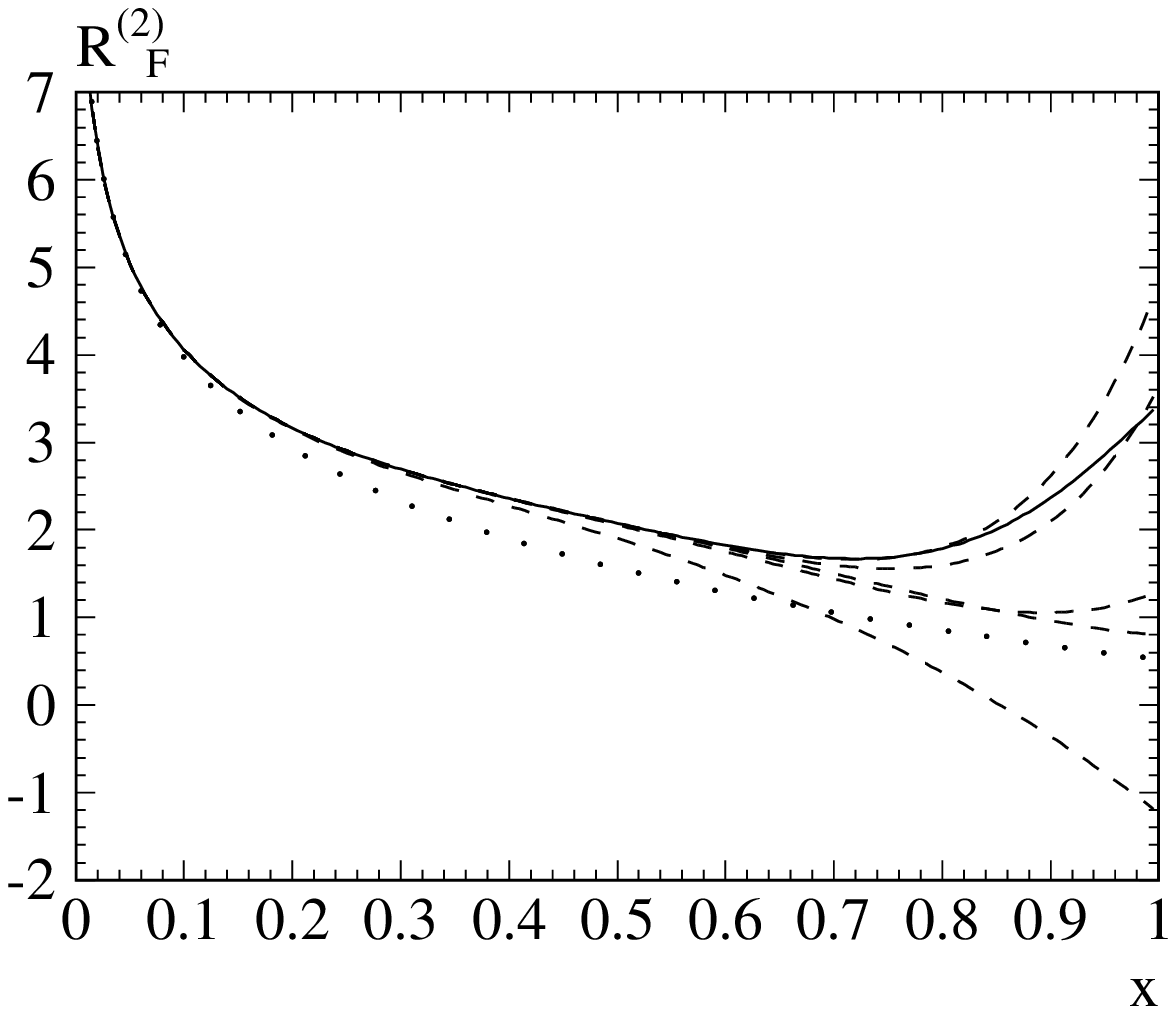}
\end{tabular}
\parbox{14.cm}{\small\bf
    \caption[]{\label{rvx.ps}\sloppy\rm
      The abelian contribution $R_A^{(2)}$, 
      the non-abelian piece $R_{\it NA}^{(2)}$,
      the contribution from light internal quark loops $R_l^{(2)}$ 
      and the contribution $R_F^{(2)}$ from the double-bubble diagram 
      with the heavy fermion in both the inner and outer loop
      as functions of $x = 2m/\sqrt{s}$.
      Wide dots: no mass terms; 
      dashed lines: including mass terms $(m^2/s)^n$ up to $n=5$; 
      solid line: including mass terms up to $(m^2/s)^6$;
      narrow dots: semi-analytical result (except for $R_F^{(2)}$). The
      scale $\mu^2 = m^2$ has been adopted.
      }}
\end{center}
\end{figure}

The results of the expansion can now be compared with those
obtained via a semi-analytical procedure using the
method of conformal mapping and Pad\'e approximation
\cite{CheKueSte96} which are valid in the whole energy range.
To conform with the conventions of \cite{CheKueSte96}, $\mu^2 = m^2$ has
been adopted for this comparison.
In Fig.~\ref{rvx.ps} successively higher orders in 
$(m^2/s)$ are included (dashed lines) and compared with 
the Pad\'e result (narrow dots).
By comparing the quadratic approximation with the one
containing also terms of order $(m^2/s)^6$ (solid line)
one observes that the quality of the approximation improves considerably.

\begin{figure}
\begin{center}
\begin{tabular}{cc}
    \leavevmode
    \epsfxsize=5.5cm
    \epsffile[110 265 465 560]{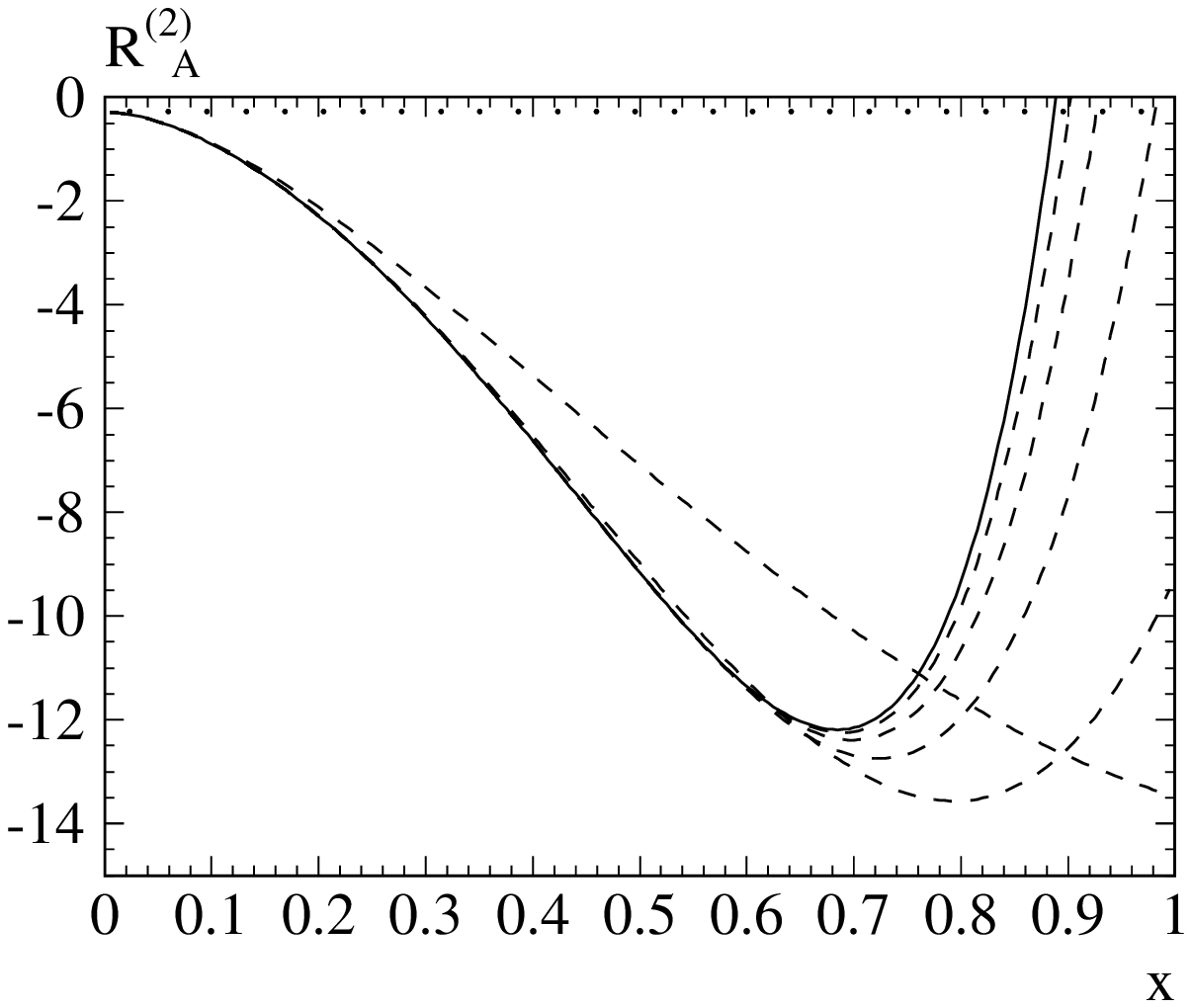} &
    \epsfxsize=5.5cm
    \epsffile[110 265 465 560]{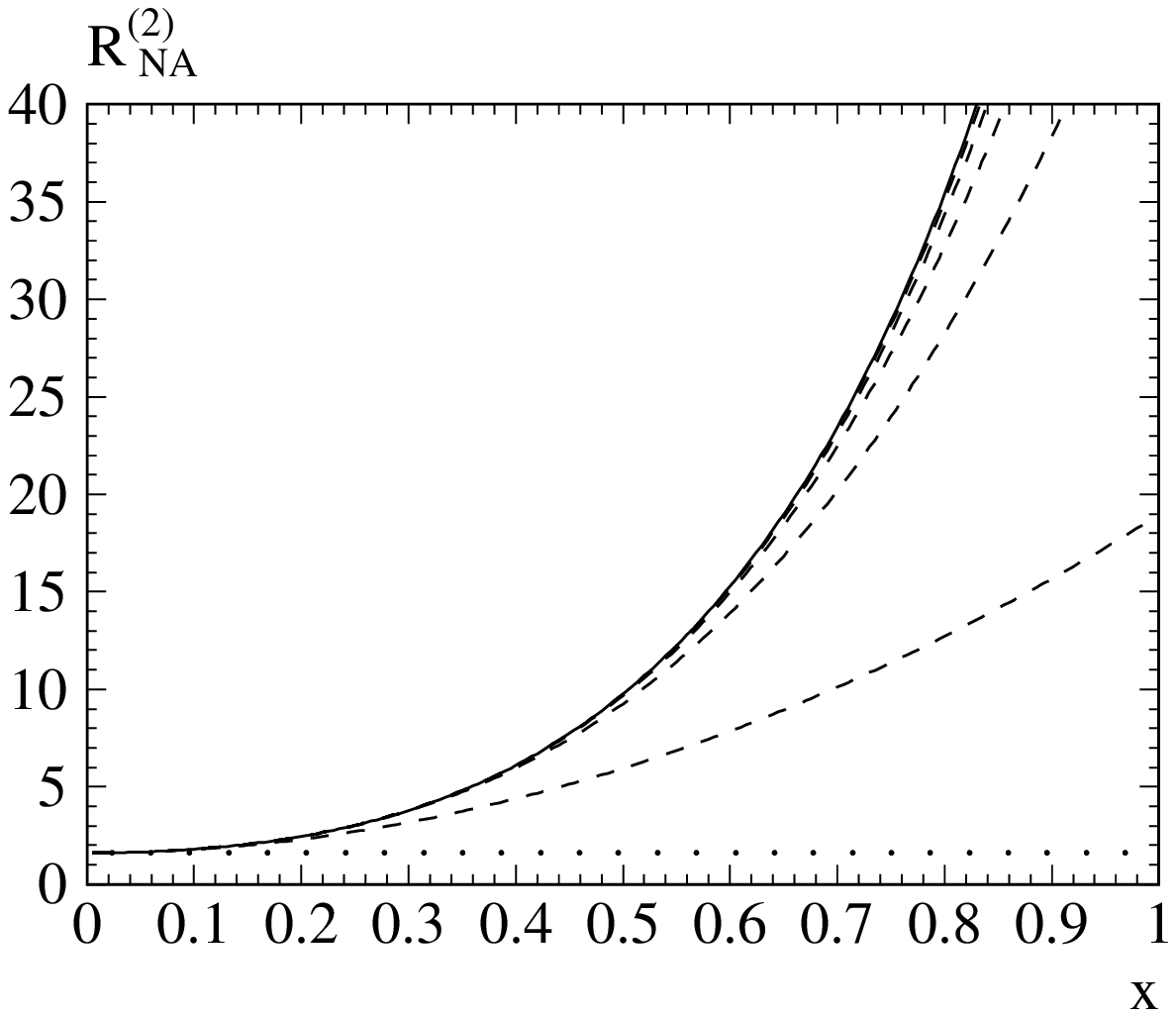}\\
    \epsfxsize=5.5cm
    \epsffile[110 265 465 560]{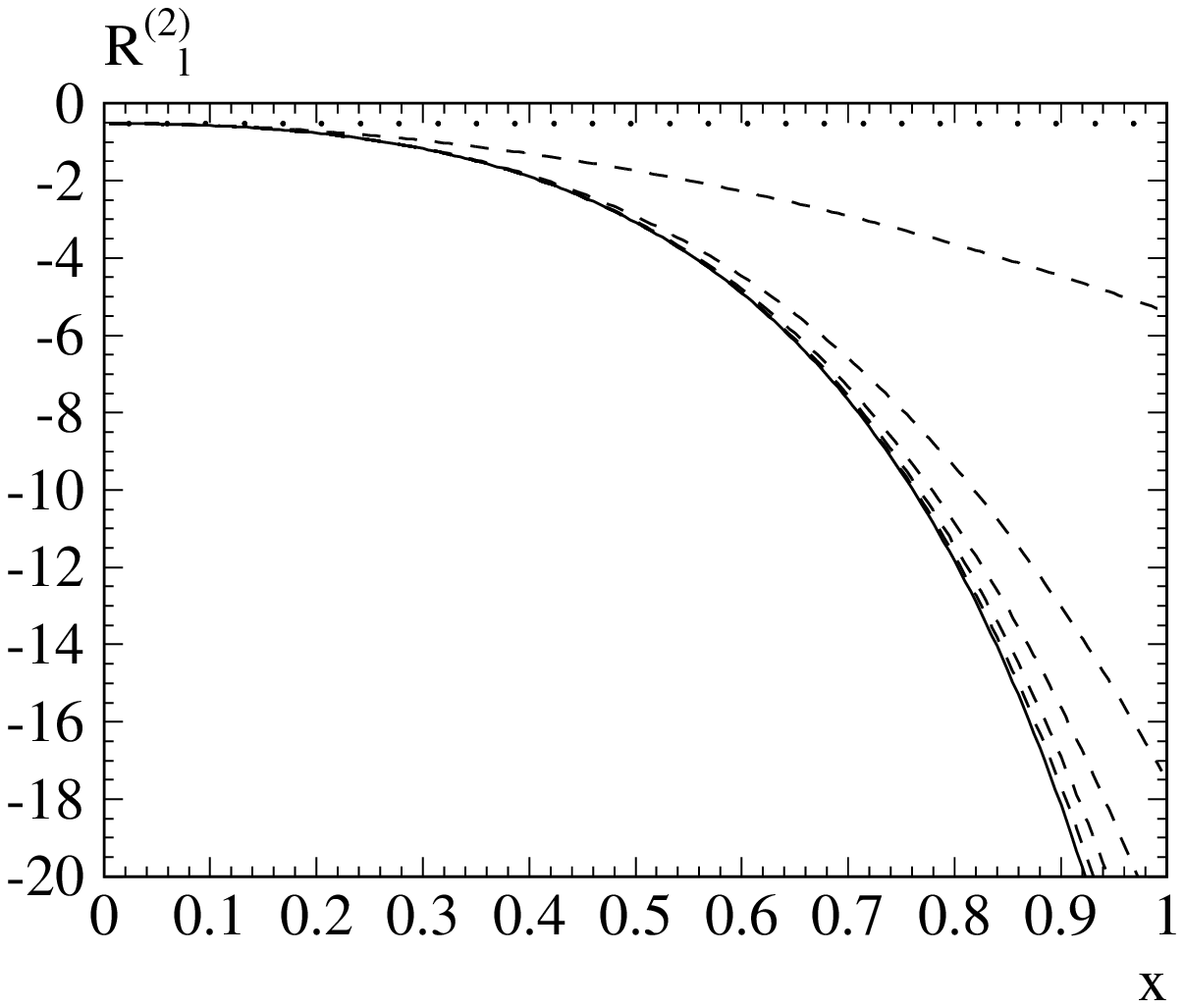} &
    \epsfxsize=5.5cm
    \epsffile[110 265 465 560]{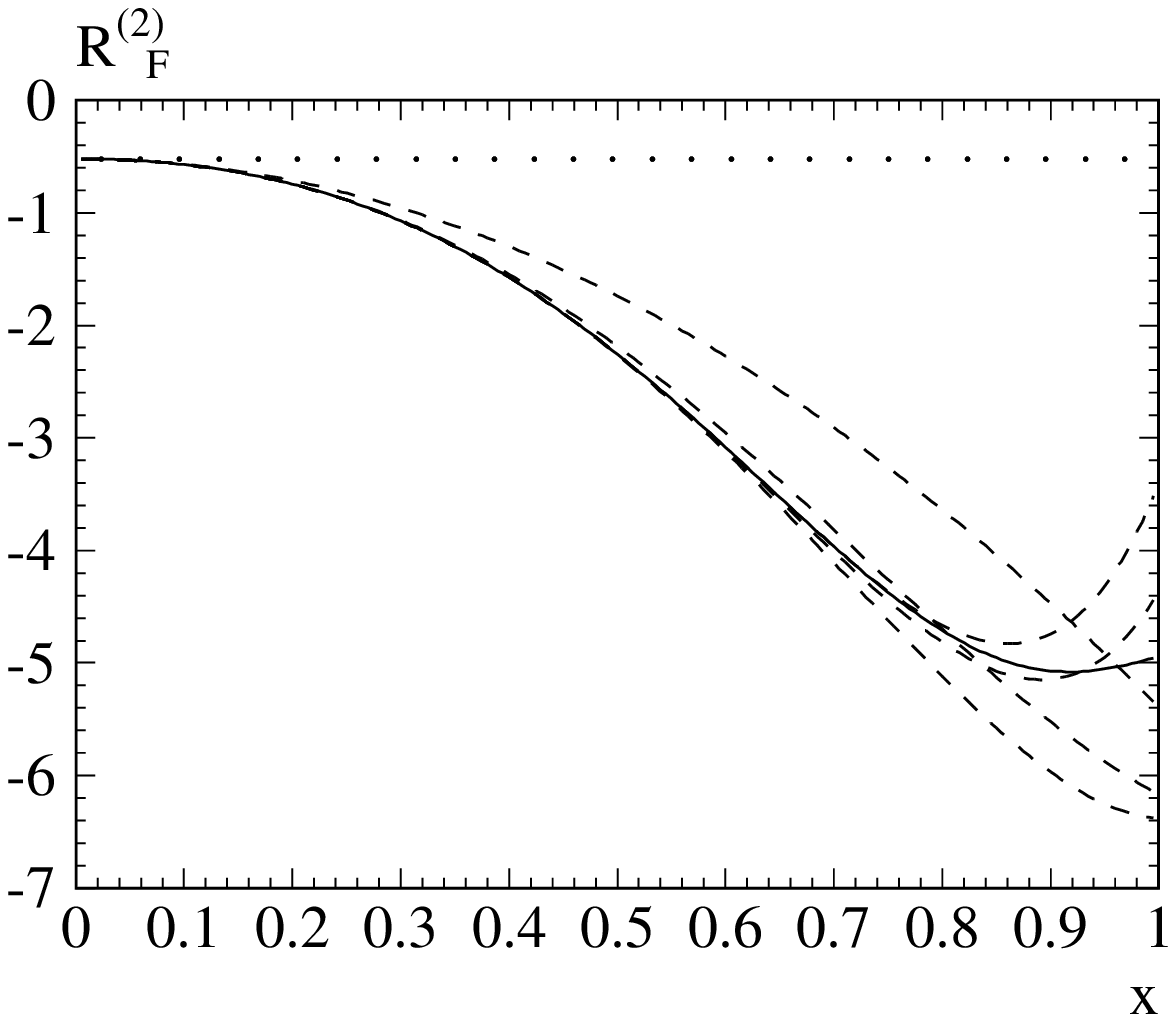}
\end{tabular}
\parbox{14.cm}{\small\bf
  \caption[]{\label{figsmus}\sloppy\rm
    $R_i^{(2)}, i=A,{\it NA},l,F$ for $\mu^2 = s$ including
    successively higher orders in $m^2/s$. The same notation as in
    Fig.~\ref{rvx.ps} is adopted. The semi-analytical result is not shown.
    }}
\end{center}
\end{figure}

For completeness we also present in Figs.~\ref{figsmus} the results for
the scale $\mu^2=s$. This choice is more adequate to the high energy
region and all functions $R_i, i=A,{\it NA},l,F$ approach the constants to be
read off easily from Eqs.(\ref{r2aos})-(\ref{r2fos}).

\begin{figure}
  \begin{center}
    \leavevmode
    \epsfxsize=8.cm
    \epsffile[110 265 465 560]{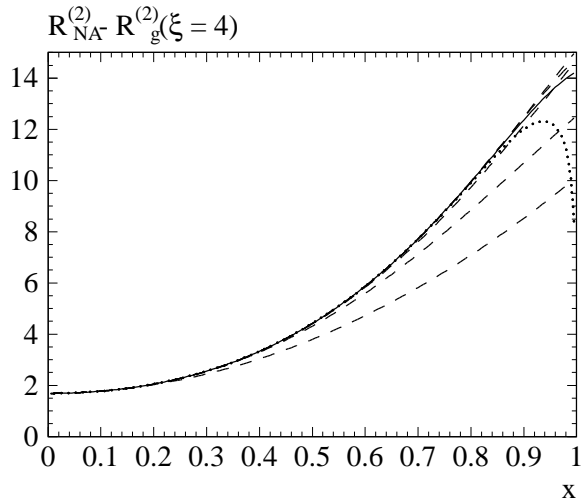}
    \hfill
    \parbox{14.cm}{\small\bf
    \caption[]{\label{figxi4}\sloppy\rm
      $R_{\it NA}^{(2)} - R_{g}^{(2)}(\xi = 4)$ over $x = 2m/\sqrt{s}$. 
      Dotted: semi-analytical result; dashed: mass terms up to $(m^2/s)^5$;
      solid: mass terms up to $(m^2/s)^6$.
      }}
  \end{center}
\end{figure}
\begin{figure}
  \begin{center}
    \leavevmode
    \epsfxsize=8.cm
    \epsffile[110 265 465 560]{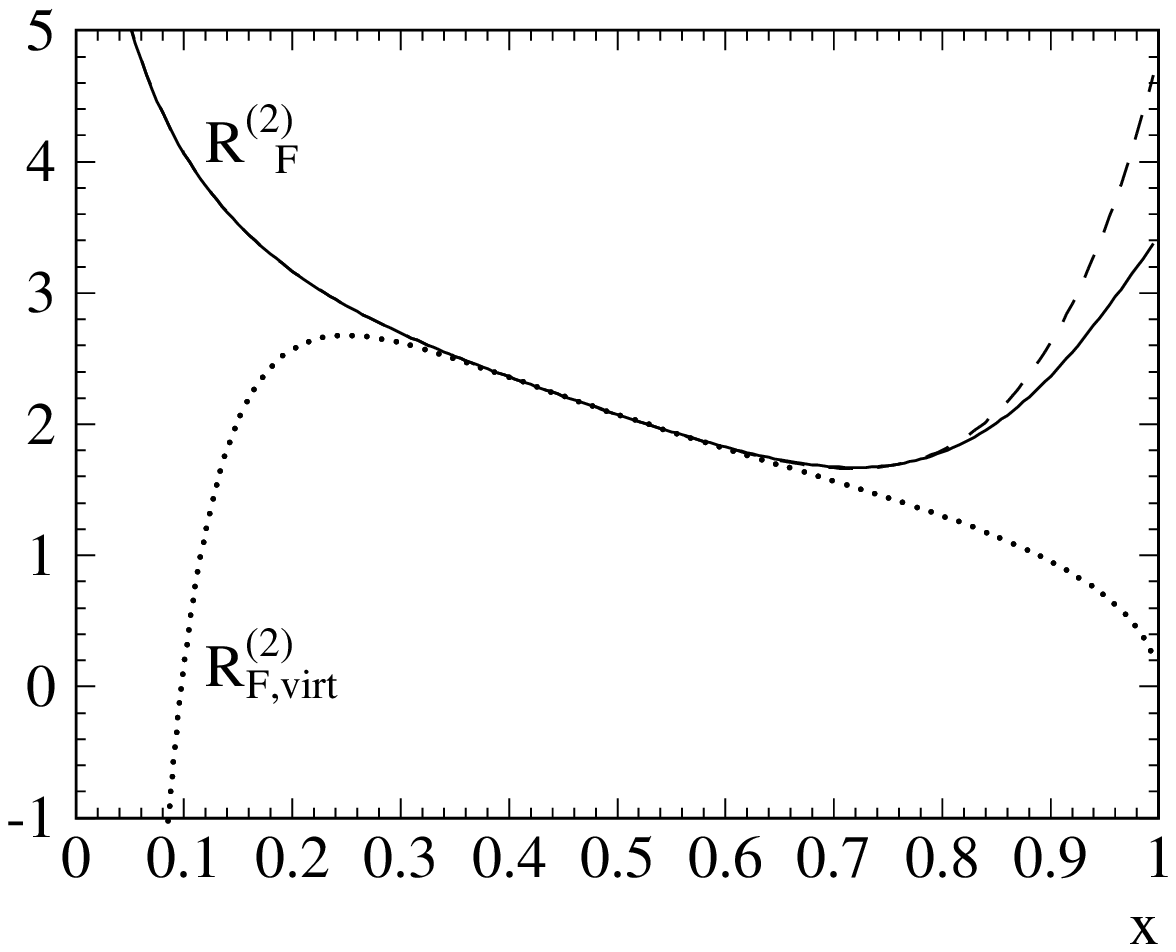}
    \hfill
    \parbox{14.cm}{\small\bf
    \caption[]{\label{rfvirt.ps}\sloppy\rm
      Comparison between the analytical result without 4-particle
      contribution ($R^{(2)}_{F,virt}$, dotted line) 
      and approximate result with terms up to
      order $(m^2/s)^5$ (dashed) and $(m^2/s)^6$ (solid line).
      }}
  \end{center}
\end{figure}
%


Let us discuss separately the high energy and the low energy regions.
For all three functions $R_{A}$, $R_{\it NA}$ and $R_l$ and values between
$x=0$ and $x=0.6$ ($x=2m/\sqrt{s}$)
the expansions including terms of order $(m^2/s)^3$ (or
more) are in perfect agreement with the semi-analytical result.
Conversely this provides a completely independent test of the method of
\cite{CheKueSte96} which did rely mainly on low energy information. Including
more terms in the expansion, one obtains an improved
approximation even in the low energy region. However, the quality of the
``convergence'' is significantly better for $R_l$ and $R_{\it NA}$
than for $R_A$. Two reasons may be responsible for this difference: 
{\rm (i)} In a high energy expansion it is presumably more difficult to
approximate the $1/v$ Coulomb singularity in $R_A$ than the mild $\ln
v$ singularity in $R_{\it NA}$ and $R_l$.
{\rm (ii)} The function $R_l$ can be approximated in the whole energy region
$2m<\sqrt{s}<\infty$ by an increasing number of terms with arbitrary
accuracy. This is evident from the known analytical form of $R_l$, a
consequence of the absence of thresholds above $2m$ in this piece. In
contrast the functions $R_A$ and $R_{\it NA}$ exhibit a four particle
threshold at $\sqrt{s} = 4m$. The high energy expansion is, therefore,
not expected to converge to the correct answer in the interval between
$2m$ and $4m$. For $R_{\it NA}$ this feature can be studied in more detail
by separating $R_{\it NA}$ into the gluonic double-bubble terms in the $\xi
= 4$ gauge \cite{CheHoaKueSteTeu96} and a
remainder. This separation is possible both for the semi-analytical
result and the expansion (Fig.~\ref{figxi4}).
The remainder approaches a constant both for $x\to 0$ and $x\to 1$. For
$0<x<0.5$ the agreement is perfect. It extends even up to $x\approx
0.9$, a fact which is quite remarkable and surprising.

For $R_F$ the expansion is also shown in Fig.~\ref{rvx.ps}. Again one observes
quick convergence for $x$ between $0$ and $0.5$. 
No (semi-)analytical result is available for the comparison with $R_F$.
However, in the region below the four particle threshold an analytical
result is available, based on the calculation of the form factor in
\cite{HoaKueTeu95}. The four fermion contribution is expected to be
small for energies just above $4m$, corresponding to $x$ just below
$x=0.5$.
Reasonable agreement between the two approaches is therefore expected in
the region around $x = 0.5$. This is indeed observed in Fig.~\ref{rfvirt.ps}.

\section{\label{seccon} Summary}

An algorithm has been described which produces the asymptotic expansion of
the three-loop vacuum polarization automatically. It generates the
relevant subdiagrams and assigns the diagrams automatically to the
programs MINCER and MATAD which evaluate the resulting massless
propagator and massive tadpole integrals. The output is compared to the
quartic terms obtained in \cite{CheKue94} and to the semi-analytical
results of \cite{CheKueSte96}, confirming both these earlier results and
the validity of the expansion down to fairly low energy values.



\end{document}